\documentclass[aps,pra,twocolumn,reprint]{revtex4}
\usepackage[english]{babel}
\usepackage[utf8]{inputenc}
\usepackage{amsmath}
\usepackage{amssymb}
\usepackage{amsthm}
\usepackage{mathtools}
\usepackage{dsfont}
\usepackage{graphicx}
\usepackage{soul}
\usepackage[normalem]{ulem}

\usepackage[draft,inline,nomargin]{fixme} 
\fxsetup{theme=color}
\usepackage{xcolor}

\newcommand{\ket}[1]{\left|{#1}\right\rangle}

\newcommand{\braket}[2]{\langle{#1}|{#2}\rangle}

\begin{document}
\title{Sequential Analysis of a finite number of Coherent states}
%\date{\today}
\author{Esteban Mart\'inez Vargas}
\email{estebanmv@protonmail.com}
%Esteban.Martinez@uab.cat}
\affiliation{F\'isica Te\`orica: Informaci\'o i Fen\`omens Qu\`antics, Departament de F\'isica, Universitat Aut\`onoma de Barcelona, 08193 Bellatera (Barcelona) Spain}

\begin{abstract}
We investigate an advantage for information processing of ordering a set of 
states over making a global quantum processing with a fixed number of 
copies of coherent states. Suppose Alice
has $N$ copies of one of two quantum states $\sigma_0$ or $\sigma_1$ 
and she gives these states to Bob. Using the optimal sequential test, the SPRT,
we ask if processing the states in batches of size $l$ is advantageous to 
optimally distinguish the two hypotheses. We find that for the symmetric
case $\{\ket{\gamma},\ket{-\gamma}\}$ there is no advantage of taking any batch size $l$. 
We give an expression for the optimal batch size $l_\text{opt}$ in the assymetric case.
We give bounds $l_\text{min}$ and $l_\text{max}$ for when $P_S\approx 1$.
\end{abstract}
\maketitle
%}}}
\section{Introduction}%{{{
The efficient detection of quantum phenomena is a matter of fundamental and 
practical importance, useful to test a fundamental theory
or create a precise detector for a technological application, for example. This kind of problem can be framed within the study of hypothesis testing \cite{CoverThomasElements2006}.
This topic can be generalized into quantum hypothesis testing \cite{StrongConverseOgawa2005,ReinforcementLBrands2020}.  
If the information is stored in one of several quantum states the problem is usually
called quantum state discrimination \cite{BarnettCroke09QSD} as the task concerns with differentiating
these quantum states. Optical systems are very relevant to quantum
technologies \cite{ClassicalCapacGiovan2004,GaussianQuantuWeedbr2012} and therefore the discrimination of optical quantum states
is an important topic of study \cite{QuantumChernofCalsam2008}.

Efficient detection implies the best use of the available resources for the discovery 
of an event in a signal. A usual approach to analyze the efficiency of a protocol
is to fix a number $N$ of resources and find the apparatus that minimizes the errors
\cite{helstrom1976quantum}.
However, in practice, it is useful to consider online, on-the-fly detectors such as
change-point detection
\cite{tartakovsky2014sequential,QuantumChangeSentis2016,ExactIdentificSentis2017,CertifiedAnsweVargas2019}.

In its simplest form, quantum state discrimination consists in 
being given a state $\rho$ with the promise of being one of two possible states:
$\rho=\sigma_0$ or $\rho=\sigma_1$ (hypothesis 0 and 1 respectively) and construct 
a quantum measurement that distinguishes them with 
the lowest possible average error~\cite{Bae2017QSDApps,BarnettCroke09QSD,Chefles2000QSD}.
We call Type-I error for guessing hypothesis 1 as true while it is false
and Type-II for guessing hypothesis 0 as true while being false.
Such measurement is described by a Positive Operator Valued Measure (POVM).
One can consider $N$ copies of states and form tensor states $\rho^{\otimes N}$.
The probability of success will be higher with more copies, as more resources are 
available \cite{DiscriminatingAudena2007}.
Then, the problem changes to distinguish between two hypotheses with the lowest possible average error using
the given number of copies. 
However, the POVM might imply highly entangled operators which can be hard to build. It is relevant
and not trivial to know how well different strategies behave with respect to
the total number of resources $N$. 

Sequential analysis is a statistical framework that addresses the issue
of optimal resource handling~\cite{wald1973sequential}.
In this framework, the desired error bounds on the Type-I and Type-II errors are fixed beforehand and the number of 
average samples needed to decide within these bounds.
The protocol that minimizes this average number of resources is called the
Sequential Probability Ratio Test (SPRT)~\cite{Wald1948OptimumChar}.

Recently the framework of Sequential Analysis has been introduced to quantum theory~\cite{QSeqAnMV2021}. 
It considers the problem of having access to quantum measurements of states.
The bounds given in Ref. \cite{QSeqAnMV2021} give us the minimum number of resources needed
using quantum measurements. It was found afterward in \cite{OptimalAdaptivLiYo2022} that
the general bounds are attainable with 
adaptive measurements. The present work can be regarded as
an extension of the sequential analysis program when considering coherent states, which
imply an infinite dimensional Hilbert space. However, here we consider a fixed (or non-adaptive)
protocol. The problem we treat here uses the SPRT and asks for the probability that 
the protocol stops with $N$ copies or less with the probabilities of 
Type-I and Type-II errors being less than or equal to given probabilities 
$\alpha$ and $\beta$ respectively. 

Three different strategies are relevant to this work.
First, we have the \emph{general} case when all the $N$
copies are available at once. This case includes possibly entangled operators
for measurement. Then when the states are available one by one,
we have the \emph{online} scenario, which implies that the protocol ignores if there
is a horizon in the number of copies and therefore is optimal
at each step of the process (this protocol is well described in \cite{Sentis2022online}).
Finally, there is the \emph{sequential} scenario, that uses the SPRT and is closely 
related to the online one. A relevant difference is that the SPRT is a test that
minimizes the average number of resources needed.

\begin{figure}
    \includegraphics[width=0.49\textwidth]{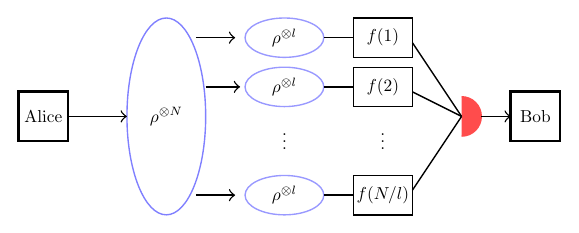}
    \caption{Bob slices the set $\rho^{\otimes N}$ into $N/l$ batches 
        $\rho^{\otimes l}$ and processes each batch in an ordering given
        by a function $f$.}
    \label{fig:bobslice}
\end{figure}

In this article, we explore the freedom of using collective quantum strategies on subsets of copies
of coherent states. The collective strategy involves an accumulation of information into
one mode \cite{UnambiguousComSedlak2008}.
%, a process used for detection of entanglement in optical systems\cite{kok_lovett_2010}.
We have a setting as in Fig. (\ref{fig:bobslice}), Alice gives a state $\rho^{\otimes N}$
to Bob
and we investigate if slicing this set into $N/l$ batches of the state $\rho^{\otimes l}$
and measuring them in an ordering given by a function $f$ 
is beneficial for Bob in terms of distinguishing which state he was given: $\sigma_0^{\otimes N}$ or 
$\sigma_1^{\otimes N}$. The function $f$ only represents the fact that we are using a
statistical method: the SPRT.

Intuition indicates that there should be a trade-off, as measurements
with more copies yield less error, however, if we make the batches too
large we will run out of copies for the SPRT, as Alice handles a finite
number of copies $N$. Therefore, given $N$ there must be an optimal batch 
size $l$ in terms of the probability of successfully identifying the given state.
We find that this is not always the case, as there are relevant instances where all
values of $l$ are equivalent. 

We first revisit the pure qubit case with unambiguous from Ref. \cite{QSeqAnMV2021}
in Sec. \ref{sec:unambqubits}.
Afterward, we treat the problem with coherent states. In Sec. \ref{sec:clasSPRT} we introduce basic notions
of the SPRT. In section \ref{sec:testGauss} we explore the problem of testing Gaussian distributions
and calculate the probabilities for the SPRT to stopping with $N_0$ copies or less depending
on which hypothesis is true, given bounds on the Type-I and Type-II errors. 
Then we introduce in Secs. \ref{sec:wignerf} and \ref{sec:accum}
the problem of measuring coherent states and the quantum strategy of accumulating the
information of several copies into one mode. This leads to the results of Sec. \ref{sec:optl}
where we explore the optimality of $l$ in several cases. We end the article in Sec.
\ref{sec:conclusions} with the conclusions.
%}}}
\section{Sequential Unambiguous POVM}%{{{
\label{sec:unambqubits}
In some cases, nonorthogonal states can be exactly distinguished if we allow the possibility of outcomes that 
don't give information. Such discrimination protocols are called unambiguous \cite{BarnettCroke09QSD}. Here
we study an unambiguous protocol for distinguishing two pure finite-dimensional states. 
Let us denote without loss of generality, the two possible states as
$\ket{\psi_0}$ and $\ket{\psi_1}$ as~\cite{MultipleCopyTAcin2005}
\begin{equation}
    \ket{\psi_a}=\cos\theta\ket{x}+(-1)^a\sin\theta\ket{y}, 
\end{equation}
where we have written them in terms of an orthonormal
basis $\ket{x}$ and $\ket{y}$ of a two-dimensional Hilbert
space and an angle $\theta\in[0,\pi/4]$ between them.
Let us denote the overlap between them as $\braket{\psi_0}{\psi_1}=\cos2\theta=:c$.
We use a three-outcome POVM because the protocol considered here is unambiguous~\cite{BarnettCroke09QSD}.
Following \cite{QSeqAnMV2021} we have the sequential probability of success
for unambiguously discriminating 2 hypotheses when $N$ copies are available goes as 
$P_S^{UA}=1-c^N$. Remarkably, this is a result that applies to a global strategy
as well as for online strategies. 
This equivalence implies that all batch sizes are equivalent. To see this last statement imagine that Bob makes
batches of size $l$ from the original set of $N$ states. 
We would therefore have the states $\ket{\Psi_a^l}=\ket{\psi_a}^{\otimes l}$.
The effective overlap between the redefined copies is $C=\braket{\Psi_0^l}{\Psi_1^l}=\braket{\psi_0}{\psi_1}^l=c^l$.
We would therefore
have $N/l$ batches. As we have batches of size $l$ then we can see this fact as
a redefinition of a copy. We have therefore the probability of success
for unambiguous discrimination of these batches as
\begin{equation}
    P_S^{UA}=1-C^{N/l}=1-c^N.
    \label{eq:psunam}
\end{equation}
The reason for the simple substitution on Eq. (\ref{eq:psunam}) is that the global 
performance of the unambiguous protocol is achieved by an online strategy \cite{QSeqAnMV2021}.
The online strategy is to apply an unambiguous POVM for each available copy.
Being an unambiguous measurement then the probability of success with $N$ 
copies coincide with the probability of
stopping at step $N$ because this measurement yields a zero error answer. Only if we get
an inconclusive outcome we would have to keep on measuring.
However, we can wait to have all the $N$ copies and make a global unambiguous measurement
and have a result with the same success probability,
therefore there is no gain in the ordering strategy by Bob in the unambiguous protocol.

A drawback of using an unambiguous protocol is that despite that it yields a no-error answer, the whole 
protocol has, in general, a lower probability of success than a two-outcome POVM. The reason
for this is that is a very restrictive protocol. Also, for mixed states, unambiguous discrimination is
possible only in very restrictive cases. 
%}}}
\section{Probability that the SPRT stops with $N_0$ samples or less}%{{{
\subsection{Classical SPRT}%{{{
\label{sec:clasSPRT}
We first review some basic notions of the SPRT theory by Wald \cite{wald1973sequential}.
Consider that we have $N_0<\infty$ independent and identically distributed (i.i.d.)
samples $x_i$ of a random variable $X$ that follows the probability distributions
$p(x|0)$ or $p(x|1)$. The $\{0,1\}$ index denotes the hypotheses $0$ or $1$
respectively.
Observe that $N_0\neq N$, $N$ will return afterward. 
We can define a useful variable
\begin{equation}
    z(x)=\log\frac{p(x|0)}{p(x|1)}
\end{equation}
where $\log$ denotes the natural logarithm.
Thus, with a set of outputs $\{x_1,\ldots,x_n\}$, we have a 
set of values $\{z(x_1),\ldots,z(x_n)\}$ that we will denote as 
$\{z_1,\ldots,z_n\}$ for simplicity. At step $n$ we define
\begin{equation}
    Z_n:=\sum_{i=1}^{n}z_i.
    \label{eq:zsumn}
\end{equation}
$Z_n$ is an example of what is known in the literature as Martingale \cite{mitzenmacher_upfal_2005},
which is a stochastic process whose mean value for step $n+1$ is the value of step $n$.
The SPRT consists in observing the value of $Z_n$ when a new sample is available. 
If $Z_n\geq C_0$ we
will accept hypothesis 0 as true. If $Z_n\leq C_1$ we will accept
hypothesis 1 as true. If $C_1 < Z_n < C_0$ continue sampling. 
It can be shown \cite{wald1973sequential} that the bounds $C_0$ and $C_1$
can be chosen such that the type I error probability is $\leq\alpha$ and
analogously, such that the type II error probability is $\leq\beta$ for given
$\alpha,~\beta\in[0,1]$.
Defining
\begin{equation}
    A:=\frac{1-\beta}{\alpha}\quad\text{and}\quad B:=\frac{\beta}{1-\alpha},
\end{equation}
we have that in a very good approximation \cite{wald1973sequential},
\begin{equation}
    C_0 \approx \log A\quad\text{and}\quad C_1\approx \log B.
\end{equation}
The SPRT is the sequential test that requires fewer samples on average \cite{Wald1948OptimumChar}.

We are given $N_0$ samples and we restrict to the SPRT.
The relevant probabilities to calculate correspond
\begin{equation}
    P_0(Z_{N_0}\geq\log A)\quad\text{and}\quad P_1(Z_{N_0}\leq\log B)
\end{equation}
where $P_i$ correspond to the probability when hypothesis $i$ is true.
Let us suppose that we are given the hypothesis 0 and 1 with equal priors therefore the total
probability of success is
\begin{equation}
    P_S = \frac{1}{2}P_0(Z_{N_0}\geq\log A)+\frac{1}{2}P_1(Z_{N_0}\leq\log B).
    \label{eq:totalprob}
\end{equation}
%}}}
\subsection{Testing Gaussians}%{{{
\label{sec:testGauss}
Suppose now that $X$ is normally distributed so that the probability
distribution when the hypothesis $i$ is true corresponds to
\begin{equation}
    p(x|i) = \frac{1}{\sqrt{2\pi}\sigma}e^{-\frac{(x-\theta_i)^2}{2\sigma^2}}.
\end{equation}

It is straightforward to show that
\begin{equation}
    z(x) = \frac{1}{2\sigma^2}(2(\theta_0-\theta_1)x+\theta_1^2-\theta_0^2).
\end{equation}

Recalling Eq. (\ref{eq:zsumn}) we have that
\begin{equation}
    \sum_{i=1}^{N_0}x_i=\frac{Z_{N_0}2\sigma^2-(\theta_1^2-\theta_0^2)}{2(\theta_0-\theta_1)}.
\end{equation}

Suppose that each $x_i$ has mean $\theta$ and variance $\sigma^2$.
Observe that $\sum_{i=1}^{N_0}x_i$ is a sum of normally distributed random
variables, therefore it is a normally distributed variable with mean $N_0\theta$
and variance $N_0\sigma^2$ \cite{lemons2002introduction}.

The stopping condition for the SPRT $Z_{N_0}\geq\log A$ translates to
\begin{equation}
    \sum_{i=1}^{N_0}x_i\geq\frac{\log A 2\sigma^2-(\theta_1^2-\theta_0^2)}{2(\theta_0-\theta_1)}.
\end{equation}
Observe that the probability that a normally distributed variable $X$ to take a value less than or equal
$\lambda$ is given by the cumulative probability $G(\lambda)$. As we want the probability that a variable 
takes a value less than or equal to some lambda we need $1-G(\lambda)$. In terms of the Error function \cite{arfken2005mathematical}
defined as
\begin{equation}
    \text{Erf}(y)=\frac{2}{\sqrt{\pi}}\int_0^ye^{-t^2}dt,
    \label{eq:erfdef}
\end{equation}
we thus have the probability
\begin{align}
    P_0(&Z_{N_0}\geq\log A)=\nonumber\\
    &\frac{1}{2}\left(1-\text{Erf}\left(\frac{2\sigma^2\log A -(\theta_1^2-\theta_0^2)-2N_0\theta_0(\theta_0-\theta_1)}{2(\theta_0-\theta_1)\sqrt{2N_0}\sigma}\right)\right).
    \label{eq:P0}
\end{align}
Analogously, we can calculate
\begin{align}
    P_1(&Z_{N_0}\leq\log B)=\nonumber\\
    &\frac{1}{2}\left(1+\text{Erf}\left(\frac{2\sigma^2\log B -(\theta_1^2-\theta_0^2)-2N_0\theta_1(\theta_0-\theta_1)}{2(\theta_0-\theta_1)\sqrt{2N_0}\sigma}\right)\right).
    \label{eq:P1}
\end{align}
    %}}}
%}}}
\section{Coherent states}%{{{
\subsection{Wigner function}%{{{
\label{sec:wignerf}
A coherent state $\ket{\gamma}$ is described by a complex number $\gamma$. In the phase
space, we can write $\gamma=q_\gamma+ip_\gamma$ with $q$ and $p$ denoting quadratures of the
electromagnetic field. The Wigner function of such a state is given by a Gaussian \cite{kok_lovett_2010}
\begin{equation}
    W_\gamma = \frac{2}{\pi}\text{exp}[-2(q-q_\gamma)^2-2(p-p_\gamma)^2].
\end{equation}
To detect a quadrature of the electromagnetic field one normally uses homodyne detection,
which allows us to detect intensity discrepancies in an electromagnetic field. Explicitly we
can detect \cite{kok_lovett_2010}
\begin{equation}
    \Delta I = \sqrt{2}|\gamma|\left\langle\frac{e^{-i\xi}\hat{a}+e^{i\xi}\hat{a}^\dagger}{\sqrt{2}}\right\rangle,
\end{equation}
for the angle $\xi$. Suppose we measure the quadrature $q$ with the coherent state $\ket{\gamma}$, which corresponds to $\xi=0$ the
probability distribution is a Gaussian with mean $\theta = q_\gamma$ and variance $\sigma^2=1/4$.
%}}}
\subsection{Multiple copies}%{{{
\label{sec:accum}
If multiple copies of coherent states are available we can accumulate the information into one
mode \cite{UnambiguousComSedlak2008}. Consider a beam splitter of transmissivity $T$ and reflexivity $R$, if the coherent 
states $\ket{\gamma}$ and $\ket{\delta}$ incide into the beam splitter it transforms to
\begin{equation}
    \ket{\gamma}\otimes\ket{\delta}\rightarrow\ket{\sqrt{T}\gamma+\sqrt{R}\delta}\otimes\ket{-\sqrt{R}\gamma+\sqrt{T}\delta}.
\end{equation}
Therefore, if $\gamma=\delta$ and we have a 50:50 beam splitter we get $\ket{\sqrt{2}\gamma}\otimes\ket{0}$.
In general, $l$ copies can be concentrated into one mode. Suppose several beam splitters are put
one after another such that they perform the unitary transformation $\ket{\gamma}^{\otimes l}\rightarrow\ket{\sqrt{l}\gamma}\otimes\ket{0}^{l-1}$ \cite{UnambiguousComSedlak2008}.
To achieve this, the beam splitters must have transmissivities and reflectivities
given by
\begin{equation}
    T_j=\frac{j}{j+1},\quad R_j=\frac{1}{j+1}.
\end{equation}
%}}}
\subsection{Optimal $l$}%{{{
\label{sec:optl}
We return to the scenario of Fig. (\ref{fig:bobslice}). Suppose that we are given $N_0=N/l$ batches
of copies of coherent states. For each batch of $l$ states, we implement the process of accumulation
from section \ref{sec:accum}. Therefore, the probability distributions we are comparing are given by 
\begin{equation}
    p(x|i) = \sqrt{\frac{2}{\pi}}e^{-2(x-\sqrt{l}\theta_i)^2},
\end{equation}
where $\theta_i$ is given by the real part of the coherent state $\theta_i=\Re(\gamma_i)$.
Notice that $\theta_i\rightarrow\sqrt{l}\theta_i$ with respect to section \ref{sec:testGauss}.
Therefore, following Eqs. (\ref{eq:P0}) and (\ref{eq:P1}) we have 
\begin{align}
    P_0(&Z_{N/l}\geq\log A)=\nonumber\\
    &\frac{1}{2}\left(1-\text{Erf}\left(\frac{\frac{1}{2}\log A -l(\theta_1^2-\theta_0^2)-2N\theta_0(\theta_0-\theta_1)}{\sqrt{2}(\theta_0-\theta_1)\sqrt{N}}\right)\right)
    \label{eq:P0Nl}
\end{align}
and 
\begin{align}
    P_1(&Z_{N/l}\leq\log B)=\nonumber\\
    &\frac{1}{2}\left(1+\text{Erf}\left(\frac{\frac{1}{2}\log B -l(\theta_1^2-\theta_0^2)-2N\theta_1(\theta_0-\theta_1)}{\sqrt{2}(\theta_0-\theta_1)\sqrt{N}}\right)\right).
    \label{eq:P1Nl}
\end{align}
Observe that $P_0(Z_{N/l}\geq\log A)$ and $P_1(Z_{N/l}\leq\log B)$ depend on $l$.

\begin{figure}
    \includegraphics[width=0.49\textwidth]{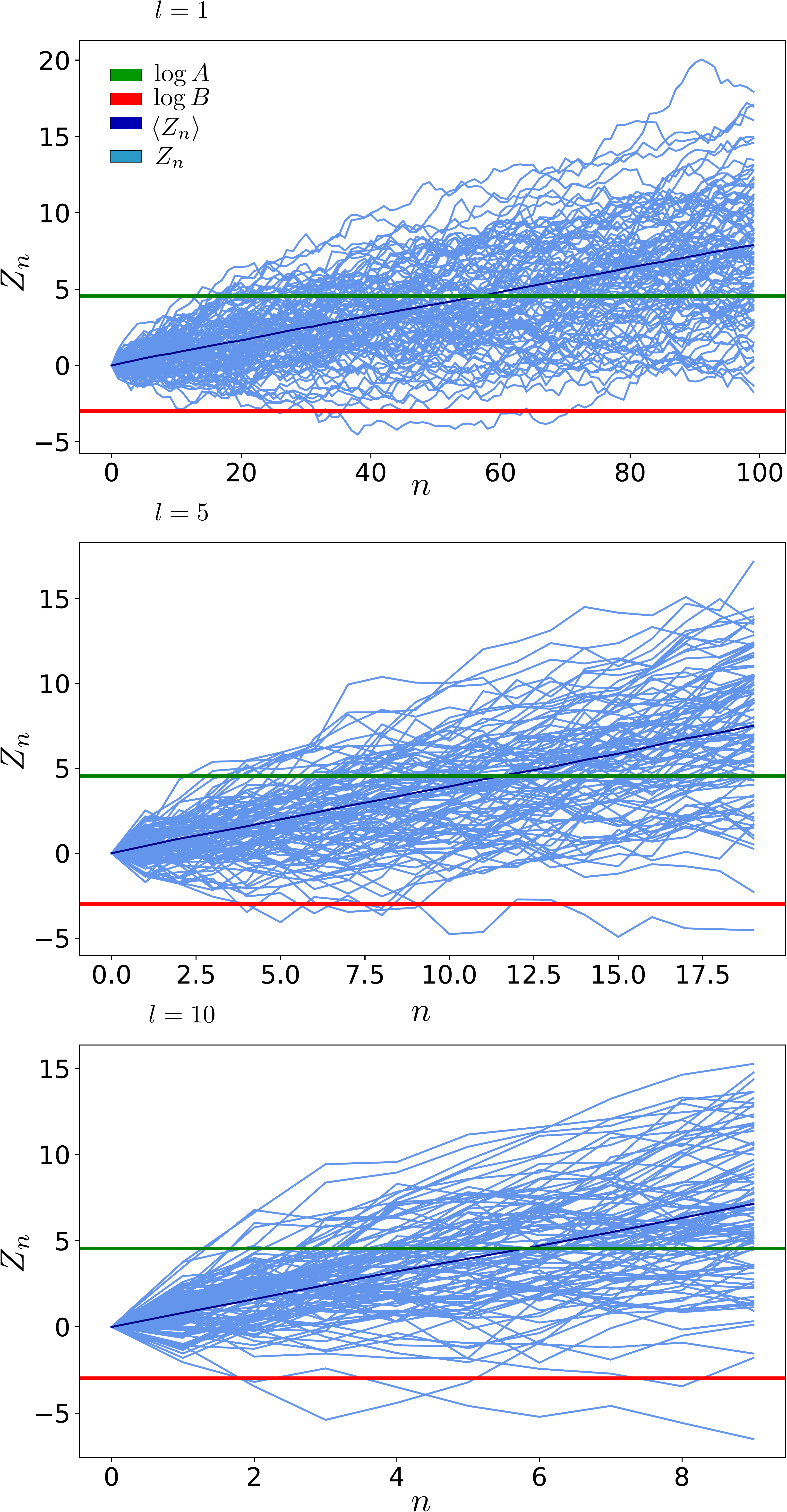}
    \caption{Examples of martingales $Z_n$ for different values of $l$ indicated above each plot. We take $1000$ samples in each figure. In each
        figure $N=100$, $\theta_0=0.1$, $\theta_1=-0.1$, $\alpha=0.01$ and $\beta=0.05$.
        The mean value of the sampling distribution is $\theta_0$.
        The pale blue paths correspond to random realizations of $Z_n$. The dark blue
        is the mean path over the $1000$ samples.}
    \label{fig:martingales}
\end{figure}

In Fig. (\ref{fig:martingales}) the SPRT is illustrated for several values of $l$. The Gaussian distribution is a numerical
approximation truncated in $\{-10,10\}$. In that figure, we observe random realizations, some of which
surpass the bound corresponding to $\log A$, which correspond to the success instances. The mean
value of the sampling distribution corresponds to $\theta_0$ and thus we see that the martingales tend to go upwards.

The cost function that needs to be optimized is the total probability from Eq. (\ref{eq:totalprob}).
It remains to optimize it over $l$. To this end, we need to investigate the sum of Error functions.
Let us then define 
\begin{align}
    y_A &:= \frac{\frac{1}{2}\log A -l(\theta_1^2-\theta_0^2)-2N\theta_0(\theta_0-\theta_1)}{\sqrt{2}(\theta_0-\theta_1)\sqrt{N}}.\\
    y_B &:= \frac{\frac{1}{2}\log B -l(\theta_1^2-\theta_0^2)-2N\theta_1(\theta_0-\theta_1)}{\sqrt{2}(\theta_0-\theta_1)\sqrt{N}}.
\end{align}
Therefore,
\begin{equation}
    P_S = \max_l\{\frac{1}{4}\left(2-\text{Erf}(y_A)+\text{Erf}(y_B)\right)\}.
\end{equation}
\subsubsection{Symmetric case}%{{{
The frequently used Dolinar receiver \cite{AnOptimumReceDolina1973} normally works with a symmetric pair of coherent states
$\{\ket{\gamma},\ket{-\gamma}\}$. If we are in this symmetric case then
we have that $\theta_0=-\theta_1$. This implies
\begin{align}
    y_A &:= \frac{\frac{1}{2}\log A -4N\theta_0^2}{2\sqrt{2}\theta_0\sqrt{N}}.\\
    y_B &:= \frac{\frac{1}{2}\log B +4N\theta_0^2}{2\sqrt{2}\theta_0\sqrt{N}}.
\end{align}
We see that there is no dependence on $l$, therefore any batch size is equally good.
%\fxnote{We see in Fig. (\ref{fig:martingales}) that we have more or less the same behavior for any
%value of $l$. The difference is associated with the asymptotic 
%convergence of probability.} 
%From this figure we observe that it is more convenient to take $l=1$ because
%more samples are available for statistical processing.
%}}}
\subsubsection{Non-symmetric case}%{{{
\begin{figure}
    \includegraphics[width=0.49\textwidth]{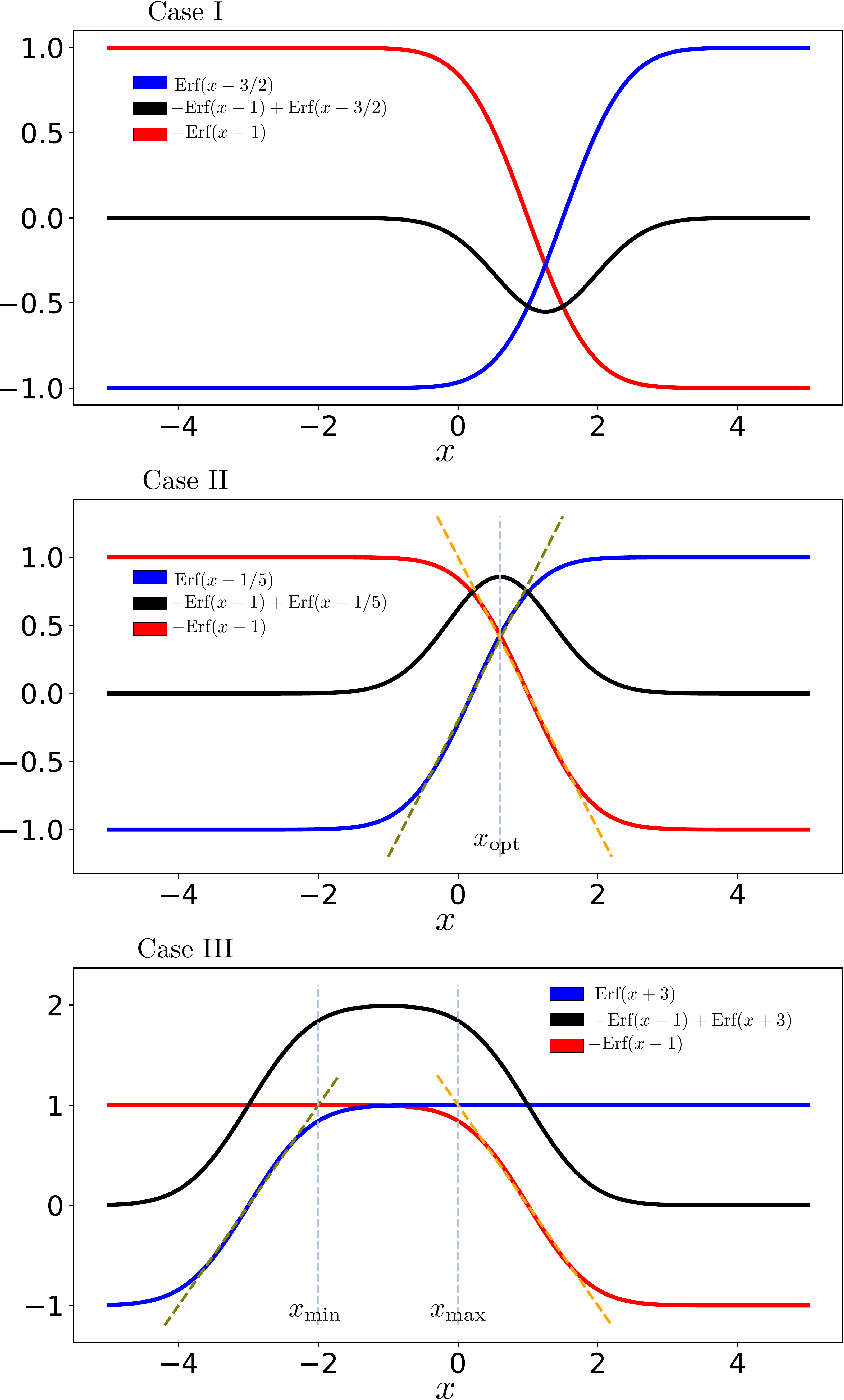}
    \caption{We plot the three possible cases for the behavior of the sum
        of Error functions that are relevant for the probability of success $P_S$.}
    \label{fig:casos}
\end{figure}

Suppose now that $\theta_0\neq-\theta_1$.
In general, $P_S$ can have three behaviors as shown in Fig. (\ref{fig:casos}).
We can change the value of $l$ such that we move in the $x$ axis of the figures in question. 
The optimization over $l$ depends on the case we have at hand.

If we are in case I there is nothing to do, we have that $P_S\leq1/2$. In this case the best
guess for the hypothesis at hand is random.

\begin{figure}
    \includegraphics[width=0.49\textwidth]{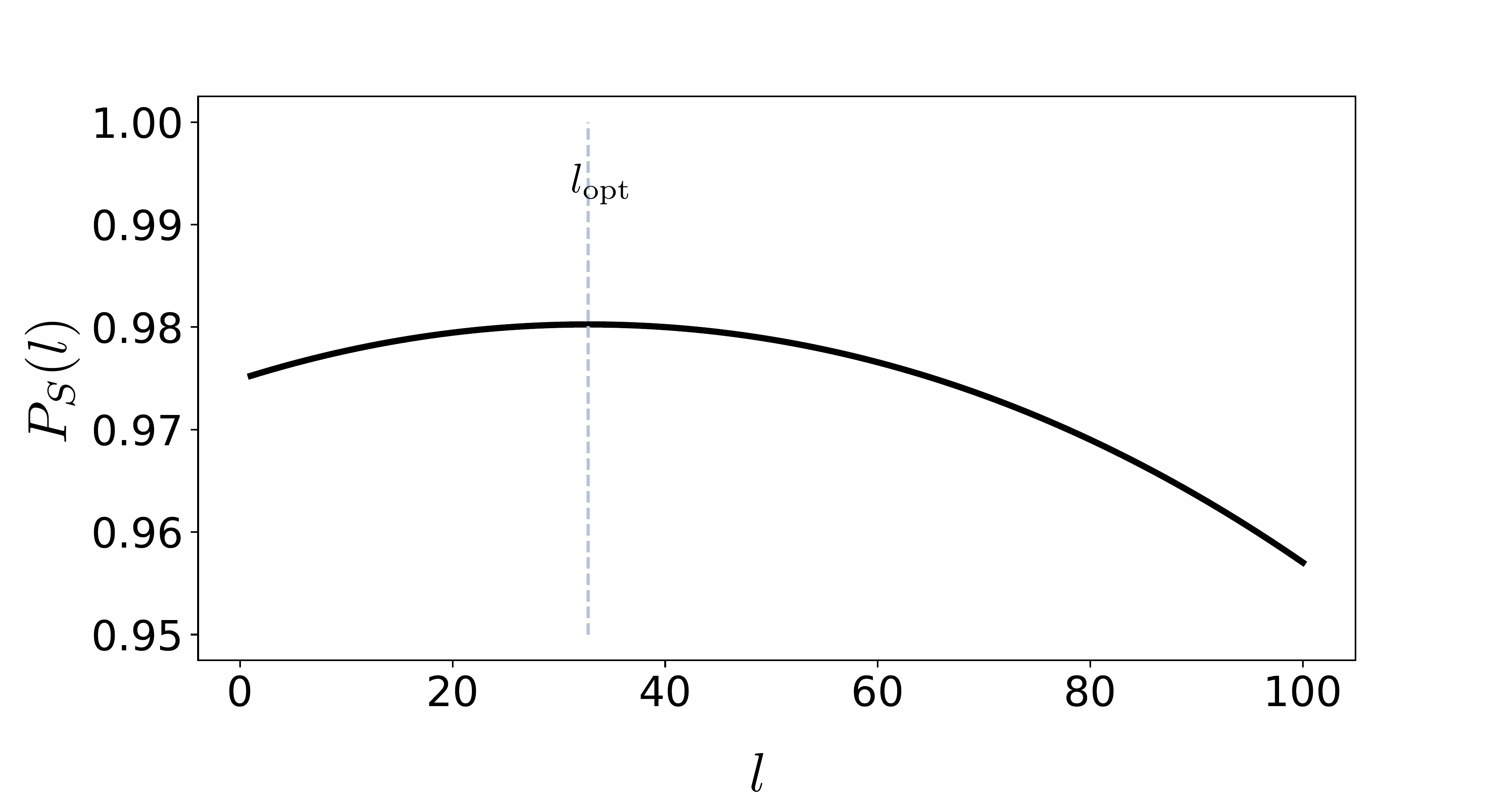}
    \caption{For this example we have $\alpha=0.00005$, $\beta=0.2$,
        $\theta_0=0.2$, $\theta_1=-0.1$ and $N=100$.}
    \label{fig:loptfig}
\end{figure}

If we are in case II observe that there is a point where the sum of Error functions attain a maximum.
This maximum can be approximated with the Taylor expansion of the exponential
around 0. Using the Eq. (\ref{eq:erfdef}) we obtain a Taylor expansion for 
the Error function around 0
\begin{equation}
    \text{Erf}(y)=\frac{2}{\sqrt{\pi}}\sum_{n=0}^\infty\frac{(-1)^ny^{2n+1}}{n!(2n+1)}.
\end{equation}
At order zero, we see that
\begin{equation}
    \text{Erf}(y)\approx\frac{2}{\sqrt{\pi}}y.
\end{equation}
Using this, by symmetry, we can obtain an approximation to the optimal value of $l$.
Notice that the point where the zero-order approximation in case II in Fig. (\ref{fig:casos})
cross each other marks the optimal value of the sum of Error functions. Therefore, the
maximum of $P_S$ is found when
\begin{equation}
    -y_A\approx y_B.
\end{equation}
We thus approximate value for the optimal $l$, we define
\begin{equation}
    l_{\text{opt}}:= N+\frac{\log A+\log B}{4(\theta_1^2-\theta_0^2)}.
    \label{eq:lopt}
\end{equation}
This value only makes sense when
\begin{equation}
    0\leq l_{\text{opt}}\leq N.
\end{equation}
In Fig. (\ref{fig:loptfig}) we have a graph of the total probability of success $P_S$
dependent on $l$ and see that it attains its maximum at $l_\text{opt}$ given by Eq. (\ref{eq:lopt}). 

If we are in case III there are limits for $x$ were $P_S\approx 1$
in Fig. (\ref{fig:casos}) as $x_\text{min}$ and $x_\text{max}$.
This implies bounds for $l$ that are defined as follows
\begin{align}
    -\frac{2}{\sqrt{\pi}}y_A&=1\nonumber\\
    \frac{2}{\sqrt{\pi}}y_B&=1.
\end{align}
These equations give the limits
\begin{align}
    l_\text{min}&:=\frac{1}{2(\theta_1+\theta_0)}\left(\frac{\log B}{\theta_1-\theta_0}+4N\theta_1+\sqrt{2N\pi}\right)\nonumber\\
    l_\text{max}&:=\frac{1}{2(\theta_1+\theta_0)}\left(\frac{\log A}{\theta_1-\theta_0}+4N\theta_0-\sqrt{2N\pi}\right).
\end{align}
These bounds are only defined for
\begin{align}
    0&\leq l_\text{min}\leq N,\nonumber\\
    0&\leq l_\text{max}\leq N.
\end{align}
%}}}
%}}}
%}}}
\section{Conclusions}%{{{
\label{sec:conclusions}
We extend the study of sequential analysis protocols for coherent states. 
Specifically, we study the probability that a specific statistical test, the SPRT 
accepts one of two possible hypotheses with $N$, a given number of resources.
In so doing we investigate the duality of collective measurements with many copies
and the necessity of having to process the measurements optimally with the SPRT. 

We find that
in the symmetric case, $\{\ket{\gamma},\ket{-\gamma}\}$ there is no advantage of taking
batches of any size. In contraposition with the adaptive protocol used in the Dolinar receiver 
\cite{AnOptimumReceDolina1973} the protocol we consider here is non-adaptive. The independence
with respect to $l$
seems to come from the fact that we are considering optimal sequential processing. 
For non-symmetric cases, two cases are relevant to us.
In the first one, there is a unique $l$ that achieves the maximum labeled $l_\text{opt}$, which is approximated
using the Taylor expansion of the Error function. The second relevant case implies a range
of values of $l$ for which, using the expansion of the Error function we define a lower bound 
$l_\text{min}$ and an upper bound $l_\text{max}$ for the range of values of $l$ that attain the optimal
$P_S$.

Operationally speaking, the SPRT shows an advantage when considering
small type-I and type-II error probabilities. However, notice that the protocol
we are considering is more general than only making a collective,
entangled measurement. The reason for this is that the batch could
be of size $N$ \emph{always} i.e. $l=N$. We show that in general, this is not
the case and that there is an advantage when taking into account 
the statistical process.

The treatment here was with the most simple quantum strategy that involves only
pure states and fixed measurements. Perhaps an adaptive strategy 
in the measurement apparatus gives more insight into when sequential information
processing is necessary \cite{OptimalAdaptivLiYo2022}.

The results from this work could be generalized to the mixed-state
case for finite-dimension states. It would be
necessary a calculation of the probabilities of Eqs. (\ref{eq:P0}) and
(\ref{eq:P1}).
%}}}
\section{Acknowledgements}%{{{
I want to acknowledge useful discussions and suggestions from 
Ramon Muñoz-Tapia, Gael Sentís and John Calsamiglia.
%}}}
\bibliography{OptimalChunksbib}
\end{document}